\documentstyle[epsf,twocolumn]{jpsj}
\newcommand{\vect}[1]{\mib #1}
\newcommand{\Eq}[1]{eq. (\ref{#1})}
\newcommand{\Fig}[1]{Fig. \ref{#1}}
\newcommand{\deffig}[5]{
\begin{figure}[tb]
  \begin{center}
    \hspace*{0em}
    \epsfxsize=8cm
    \null\ 
    \epsfbox{#4}
    \caption{#3}
    \label{#1}
  \end{center}
\end{figure}
}

\title{
Loop Algorithm for Heisenberg Models with Biquadratic Interaction
and Phase Transitions in Two Dimensions
}
\author{Kenji~HARADA$^1$ and Naoki~KAWASHIMA$^2$}
\inst{
${}^1$\ Department of Applied Analysis and Complex Dynamical Systems, 
Kyoto University, Kyoto 606-8501\\
${}^2$\ Department of Physics, Tokyo Metropolitan University, 
Tokyo 192-0397\\
}
\recdate{September 20, 2000}
\abst{
We present a new algorithm for quantum Monte Carlo simulation based on
global updating with loops.
While various theoretical predictions are confirmed in one dimension,
we find, for $S=1$ systems on a square lattice with
an antiferromagnetic biquadratic interaction,
that the intermediate phase between the
antiferromagnetic and the ferromagnetic phases is disordered and 
that the two phase transitions are both of the first order 
in contrast to the one-dimensional case.
It is strongly suggested that the transition points coincide with those
at which the algorithm changes qualitatively.
}
\kword{
loop algorithm, quantum Monte Carlo,
biquadratic interaction, Heisenberg model, phase transition, one dimension,
two dimensions
}

\begin{document}
\sloppy
\maketitle


Most theoretical studies on the model described by the Hamiltonian
\begin{equation}
  -{\cal H} = \sum_{(ij)} \left(J_L (\vect{S}_i\cdot\vect{S}_j)
                  +J_Q (\vect{S}_i\cdot\vect{S}_j)^2 \right)
\end{equation}
have been lead mainly by purely theoretical motivations.
There are many special points in the phase diagram where one
can obtain some rigorous results for the model with $S=1$ in one dimension
\cite{Uimin1970,Lai1974,Sutherland1975,Takhtajan1982,Babujian1982,Babujian1983,Barber1989,Kluemper1989,Kluemper1990}.
In particular, it is generally believed
that at each phase transition point,
the model is integrable and a Bethe ansatz solution exists.
Recent experiments on a quasi one-dimensional system,
LiVGe$_2$O$_6$\cite{Millet1999}, however,
provided us with a new motivation for studying 
this model.
The experimental results suggested the presence of a biquadratic
interaction of considerable magnitude.
The succeeding numerical work\cite{Lou2000} indicated that
the magnitude was not sufficiently large to take the system 
out of the Haldane phase.
However, it also confirmed that the biquadratic interaction considerably
affects the nature of the system.
Since there is no reason to expect a non-negligible contribution
of the biquadratic interaction only in the one-dimensional case,
it seems reasonable to start studying the system 
in higher dimensions.

Unlike the one-dimensional case, our understanding of systems in
higher dimensions is limited, due to a lack of exact solutions
and powerful field-theoretical methods.
In this paper we investigate the model with a Monte Carlo method 
using a new algorithm based on the concept of
the loop algorithm\cite{Kawashima1994}.
It is found that the phase transition points for the $S=1$ case
in two dimensions coincide with (or are located very close to)
the points where the types of graphs in the loop algorithm change
as shown below.

First, it should be noted that, similar to the conventional
world-line algorithm, the new algorithm also
encounters the negative sign difficulty with the ferromagnetic
biquadratic interaction, that is, in the case of $J_Q < 0$.
Provided that $J_Q \ge 0$, the new algorithm is available 
for the model when
\begin{equation}
  (2S^2 - 2S +1) < J_L/J_Q
  \quad\mbox{or}\quad
  J_L/J_Q < - 2 S (S-1),
\end{equation}
for general spin $S$ and in arbitrary dimensions.
Outside this region, we encounter the negative sign difficulty.
In what follows we do not discuss algorithms in such cases,
since they would not be of much use.
In contrast to the case of frustrated spin models, the difficulty
remains even if there are no closed (frustrated) cycles in the system.
We encounter serious difficulty even in one dimension.
It is found, however, that the case of $S=1$ is exceptional.
As we see below, we are exempt from negative signs
as long as $J_Q \ge 0$ holds.


The method we use here is the world-line Monte Carlo method.
Recently, a number of new techniques have been developed
for reducing the long computational correlation times that are
problematic in many applications.
Loop algorithms were found to be particularly successful\cite{Kawashima1994,Harada1998}.
A loop algorithm consists of two procedures; one for constructing
a graph (graph assignment) and the other 
for generating a configuration (loop or cluster flip).
In general, both procedures are carried out probabilistically. The graph
assignment is applied to each plaquette on nearest-neighbor sites and
short (formally infinitesimal) intervals of imaginary time.  In simple
cases, we can easily compute the graph assignment probability once we
express the Hamiltonian in the following form,
\begin{equation}
  - {\cal H}_{\rm local} = \sum_{G\in\Gamma} a(G) \hat\Delta(G)
  \quad (a(G) > 0),
  \label{eq:Decomposition}
\end{equation}
where ${\cal H}_{\rm local}$ is the local Hamiltonian
and the right-hand side is a sum of operators over a certain set of graphs $\Gamma$.
The symbol $\hat\Delta(G)$ denotes an operator whose matrix elements
are one when the graph $G$ matches the initial and the final
spin states, and zero otherwise.
The probability for assigning a graph $G$ to an interval of the length
$\Delta\tau$ is simply $a(G)\Delta\tau$ if the matrix element of
$\hat\Delta(G)$ corresponding to its current state is one.
When no graph is assigned for a plaquette, we assign the
identity graph which is simply a pair of straight lines parallel to the
time axis.  After all graph assignments, each loop which consists of
graph-linked spins is independently flipped.

In the case of $S>1/2$, we express each spin operator as a sum of
$2S$ Pauli operators\cite{Kawashima1994,Harada1998},
$$
  \vect{S}_i \Rightarrow \frac12 \sum_{\mu=1}^{2S} \vect{\sigma}_{i,\mu},
$$
rendering the Hilbert space greater than the original one.
In order to eliminate contributions from unnecessary states,
we insert a symmetrization operator $\hat P$ at $\tau = \beta$.
This operator projects the entire phase space into the original space in which
$(\sum_{\mu=1}^{2S}\sigma_{i,\mu}/2)^2 = S(S+1)$ for all $i$.
In addition, when we know that the negative signs always appear in pairs
that cancel each other,
we can simply replace negative off-diagonal elements of
the Hamiltonian by their absolute values.
Thus, in general, we can replace \Eq{eq:Decomposition} by
\begin{equation}
  [|- {\cal H}_{\rm local}|]_s = 
  \left[ \left( \sum_{G\in\Gamma} a(G) \hat\Delta(G) \right)\right]_s,
  \label{eq:DecompositionII}
\end{equation}  
where $a(G)$ are positive numbers
and $|Q|$ denotes an operator whose off-diagonal matrix elements 
are equal to the absolute values of corresponding elements of $Q$.
The symbol $[\cdots]_s$ denotes the symmetrization, that is,
$[\cdots]_s \equiv \hat P \dots \hat P$.

The following identities are useful for obtaining an expression of the
form of \Eq{eq:DecompositionII} in the present case.
\begin{eqnarray}
  [ \vect{S}_i\cdot \vect{S}_j ]_s  & = & S^2(-[ 1 ]_s + 2\hat\Delta_{\rm SC}),
  \\ {}
  [ (\vect{S}_i\cdot \vect{S}_j)^2 ]_s & = &  
  S^2( (S^2+1) [ 1]_s -2(2S^2-2S+1)\hat\Delta_{\rm SC} \nonumber \\
  & & +(2S-1)^2\hat\Delta_{\rm DC}  ).
\end{eqnarray}
The operator $\hat \Delta_{{\rm SC}}$ is defined to be 
$[\hat\Delta(G_{{\rm SC}})]_s$ with any single cross graph $G_{{\rm SC}}$
(\Fig{fig:Graphs}).
Clearly, the definition does not depend on the choice of $G_{{\rm SC}}$
because of the symmetrization.
\deffig{fig:Graphs}{3.5cm}
{
The four types of graphs. 
The vertical direction corresponds to the imaginary time.
Each pair of grey vertical represents two Pauli spins
that constitute one of the original spins, in the case of $S=1$.
}{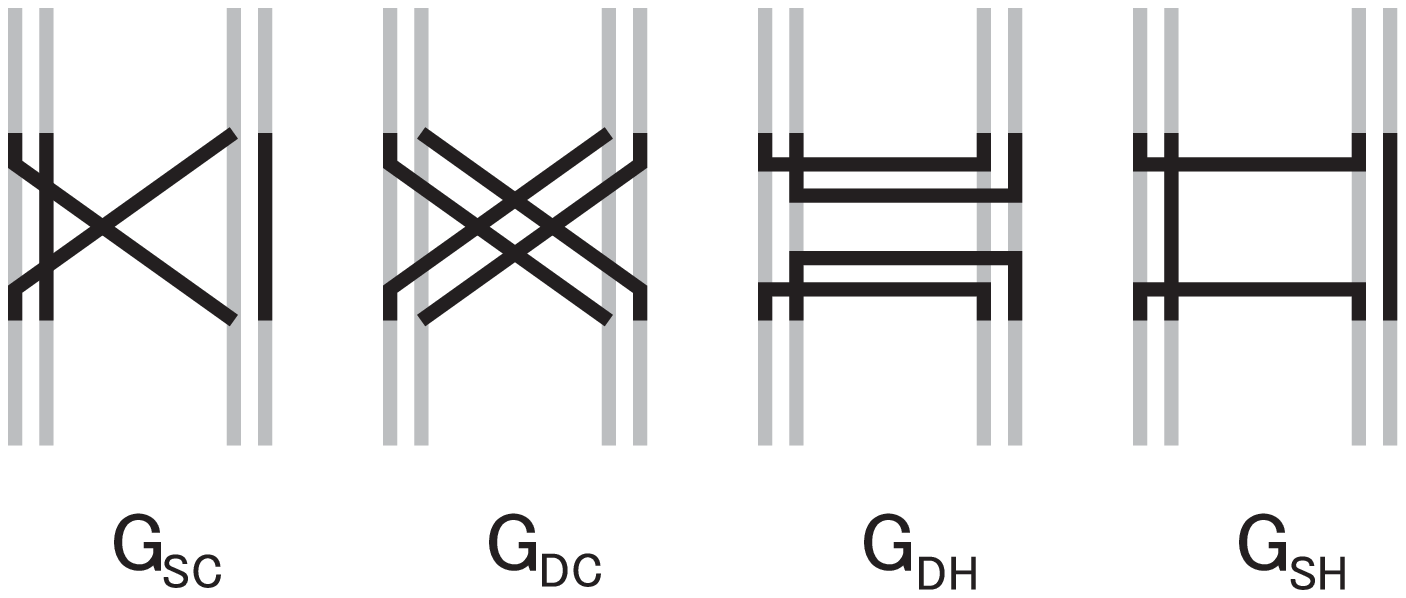}{0.4}
The other operator $\hat \Delta_{{\rm DC}}$, is defined similarly with 
a double cross graph $G_{\rm DC}$.
Then, it is easy to obtain the following expression
\begin{eqnarray}
  & & [ \left| - {\cal H}_{\rm local} \right| ]_s
  \equiv [| J_L (\vect{S}_i\cdot \vect{S}_j) 
  + J_Q (\vect{S}_i\cdot \vect{S}_j)^2 |]_s
  \nonumber \\
  & & = 
  \left[\left| 2\left(K_L - K_Q\frac{2S^2-2S+1}{(2S-1)^2}\right) 
  \hat\Delta_{\rm SC} 
  + K_Q \hat\Delta_{\rm DC} \right|\right]_s,
  \label{eq:FerromagneticCase}
\end{eqnarray}
where $K_L \equiv S^2 J_L$ and $K_Q \equiv S^2(2S-1)^2 J_Q$.
Unimportant constant terms have been omitted.
If all the coefficients are non-negative, 
this equation already has the form of \Eq{eq:DecompositionII}. 
Therefore, if $K_Q>0$ and $K_L/K_Q > (2S^2-2S+1)/(2S-1)^2$, 
\Eq{eq:FerromagneticCase} defines a valid algorithm.
This algorithm is termed the ferromagnetic algorithm in the following.

If $J_L < 0$ and $|J_L|$ is sufficiently large,
we do not have negative sign configurations.
We can see this easily using the unitary transformation
of spin operators on one of two sublattices: 
$S^x \rightarrow -S^x$, $S^y \rightarrow -S^y$ and $S^z \rightarrow S^z$.
It is equivalent to replacing $\vect{S}_i \cdot \vect{S}_j$ by
$\vect{S}_i \circ \vect{S}_j \equiv -S_i^x S_j^x -S_i^y S_j^y + S_i^z S_j^z$
which yields the following equation.
\begin{eqnarray}
  & & [|- {\cal H}_{\rm local}|]_s
  \equiv [|J_L (\vect{S}_i\circ \vect{S}_j)
  + J_Q (\vect{S}_i\circ \vect{S}_j)^2|]_s \nonumber \\
  & & =  \left[\left|
         2\left(-K_L-\frac{2S(S-1)}{(2S-1)^2} K_Q\right)\hat\Delta_{\rm SH}
        +K_Q\hat\Delta_{\rm DH}
         \right|\right]_s,
\end{eqnarray}
where $\hat\Delta_{\rm SH}$ and $\hat\Delta_{\rm DH}$ are 
the operators
corresponding to the single horizontal graph and the double horizontal graph
(\Fig{fig:Graphs}), respectively,
analogous to $\hat\Delta_{\rm SC}$ and $\hat\Delta_{\rm DC}$ mentioned above.
The coefficients are all non-negative if $K_Q>0$ and 
$K_L/K_Q \le -2S(S-1)/(2S-1)^2$.
In this case, 
the above expression is of the form of \Eq{eq:DecompositionII}
providing us with a valid algorithm.
We call this algorithm antiferromagnetic.

On the other hand, in the intermediate region 
$-2S(S-1)/(2S-1)^2 <$ $K_L/K_Q$ $< (2S^2-2S+1)/(2S-1)^2$, 
we cannot avoid the negative matrix elements in general.
However, in the case of $S=1$, we see that the negative signs
always cancel each other to give a positive global configuration.
This is because only the matrix elements between a state with two
holes (the site with $S^z=0$) and a state with 
antiparallel spins are negative, and all the others are positive.
Since the number of holes changes only in transitions corresponding
to this type of matrix element,
the number of pair creations of holes must be equal to that
of pair annihilation.
In order to construct a loop algorithm in this case,
we note that
\begin{eqnarray}
  [ |- {\cal H}_{\rm local}| ]_s &=&
  \left[ \left| 
    \left(K_Q - K_L \right) \hat\Delta_{\rm DH} 
  + K_L \hat\Delta_{\rm DC} \right| \right]_s.
\end{eqnarray}
This gives a set of probabilities required for the algorithm
called intermediate.

It is worth noting that in the case of $S=1$, both of the algorithmic
transition points correspond to special points in the physics
of the model.
Using standard parameterization, 
we define $\theta$ and $J$ so that
$K_L = - J \cos \theta$ and $K_Q = - J \sin \theta$.
Then, the algorithmic transition from the ferromagnetic algorithm 
to the intermediate algorithm occurs at $\theta = -\pi/2$, whereas
the transition from the intermediate algorithm to the antiferromagnetic one
occurs at $\theta = -3\pi/4$.
The latter ($\theta = -3\pi/4$) is known to be the real transition point
in one dimension
at which the non-magnetic (dimer) ground state switches to the ferromagnetic 
ground state.
The former ($\theta = -\pi/2$) corresponds to the point 
at which a Bethe ansatz solution
\cite{Kluemper1989,Parkinson1987}
is available.
At this particular point, the algorithm consists of only one type of graph,
i.e., the double horizontal graphs.
In terms of graphs, the system is very similar to two independent 
anti-ferromagnetic Heisenberg models with $S=1/2$.
The only difference is that the two systems are coupled with each other
through the symmetrization operator at $\tau = \beta$.
The situation at $\theta = -3\pi/4$ is similar.
In graphical terms, at $\theta=-3\pi/4$, 
the system is two $S=1/2$ Heisenberg ferromagnets 
coupled to each other only at $\tau=\beta$.
We suspect that there is a relationship between these facts and 
the integrability of the system at these points.

It should also be noted that the locations of the algorithmic transition 
points do not depend on the dimensionality or the lattice structure since 
they only depend on properties of the two-point Hamiltonian.
As we will see below, the present numerical results strongly suggest that
the algorithmic transition points are special not only in terms of the
algorithm
but also in real physics even in higher dimensions.


To evaluate the validity of the method, 
we have verified various predictions based on exact solutions
in one dimension.
We first locate the phase transition points.
It is easy to do so for the dimer-ferromagnetic phase transition
at $\theta = -3\pi/4$ since there is a very clear discontinuity 
there in the first derivative of the energy with respect to $\theta$,
indicating that the ground state switches from the dimer state to the
ferromagnetic state.
In addition, there is a clear discontinuity in the magnetization 
as a function of $\theta$ at $\theta = -3\pi/4$.

On the other hand, 
it is difficult to see a singularity in energy 
as a function of $\theta$ around $\theta = -\pi/4$.
We compute the Binder cumulant ratio
\begin{equation}
  g_Q \equiv \frac12
  \left(
  3-\frac{\langle\langle Q^4 \rangle\rangle}{\langle\langle Q^2 \rangle\rangle^2}
  \right),
\end{equation}
for the staggered magnetization and the dimer order parameter.
Here, $Q$ is the operator of an observable as given above and 
$\langle\langle Q^n \rangle\rangle$ denotes 
the temporal average as well as the thermal average, i.e.,
\begin{equation}
  \langle\langle Q^n \rangle\rangle
  \equiv \frac1{\beta^n}
  \int_0^{\beta} {\rm d}\tau_1 \cdots {\rm d}\tau_n 
  {\cal T} \langle Q(\tau_1) \cdots Q(\tau_n) \rangle,
\end{equation}
where ${\cal T}$ denotes the time-ordered product and the single bracket 
denotes the thermal average.
We have seen that, with a fixed value of $\beta/L = 1/2$ for example,
$g_{M_s}$ curves (as a function of $\theta$) for various $L$ 
intersect one another at $\theta = -\pi/4$.
In addition, when plotted as a function of $\beta$ at $\theta = -\pi/4$,
a curve for a fixed system size has a peak around $\beta \sim \beta_{\rm peak}$.
The location of the peak $\beta_{\rm peak}$ is proportional to $L$ and
the peak height is independent of the size, indicating that
$\theta = -\pi/4$ is critical and $z=1$, as expected.
We also compute the correction terms in $E(L,T=0)$ expanded with respected to $1/L^2$
and in $C(L=\infty,T)$ expanded with respect to $T$.
From the coefficients of these we have estimated the central charge and the velocity.
They are found to be consistent with the theoretical predictions $c = 1$ 
and $v = \pi/3$.


Now we turn to the two-dimensional case.
We have performed simulations for up to $L=64$ and $\beta=32$
at various values of $\theta$.
First, similar to the one-dimensional case, we can easily locate the
phase transition to the ferromagnetic phase in the energy curve
as a function of $\theta$ (\Fig{fig:EnergyII}).
\deffig{fig:EnergyII}{6cm}
{
Total energy in the case of $S=1$ in two dimensions.
The inset is a magnified view of the vicinity of the algorithmic
transition point $\theta = -\pi/2$.
The data points are for sufficiently large $L$ and $\beta$
so that they have no visible dependence on $L$ and $\beta$.
}{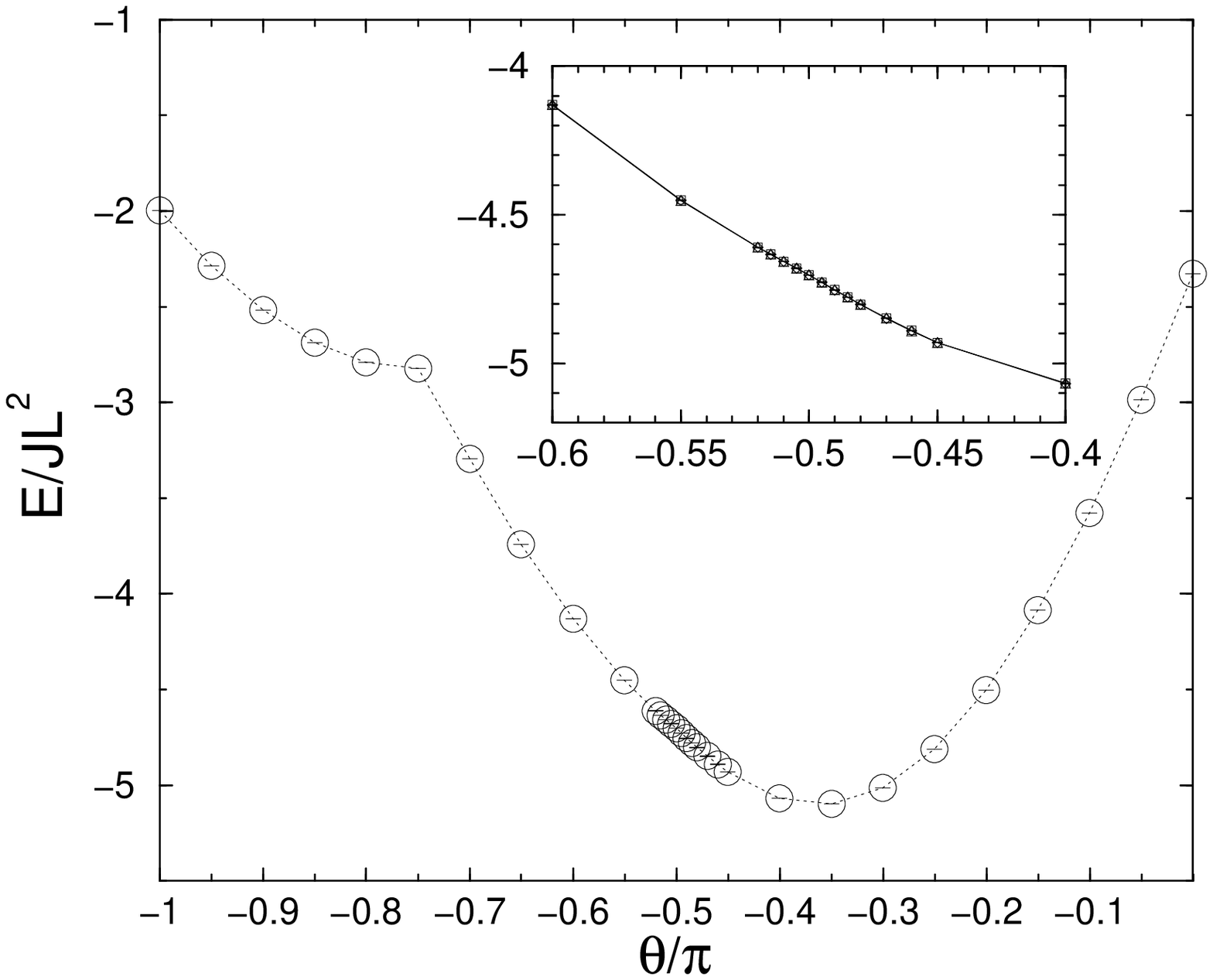}{0.4}
It is easy to prove that the ground state is purely ferromagnetic
in the region $\theta < -3\pi/4$. The present result suggests that 
as soon as the simple proof becomes invalid, the ground state switches
from the ferromagnetic state to another state 
with different symmetry, in the same way as the one-dimensional case.
We can also observe the discontinuity in the magnetization curve.
Thus, we see that one of the phase transitions coincides with one of the two
algorithmic transition points.

For the other transition point, we compute the staggered magnetization.
Since the system has a finite staggered magnetization at $\theta = 0$, 
we expect at least a transition from
the antiferromagnetic phase to another phase.

\deffig{fig:StaggeredII0500}{6cm}{
Staggered magnetization per spin
at the algorithmic transition point, $\theta = -\pi/2$.
}{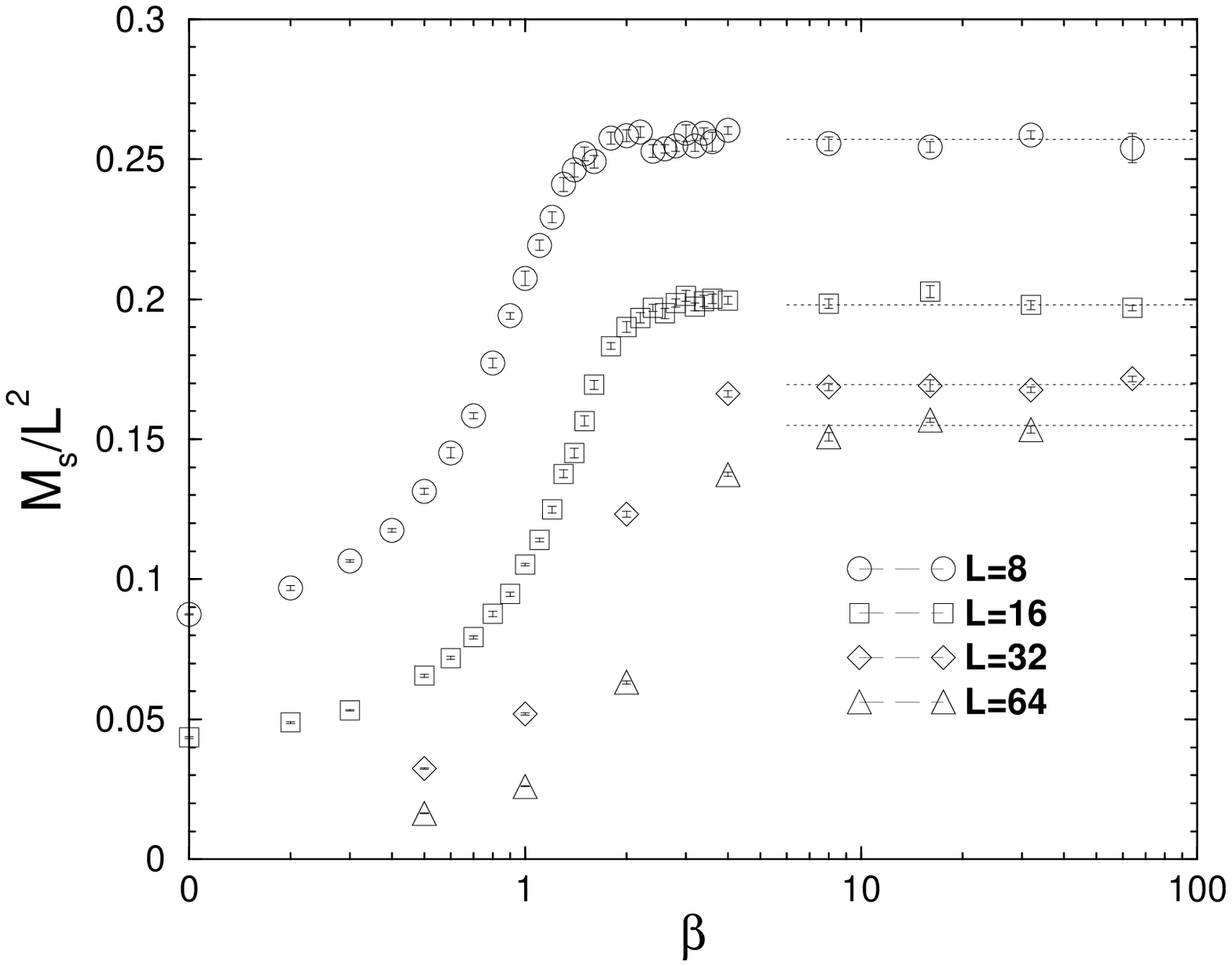}{0.4}

We observe that, for each system size, the staggered magnetization
increases initially and saturates at its zero-temperature value 
as the temperature is lowered.
As an example, the staggered magnetization for various
system sizes and temperatures is plotted at $\theta = -\pi/2$,
in \Fig{fig:StaggeredII0500}.
To the zero-temperature values for various system sizes
we fit the function 
$$
  M_s(L,\theta)/L^2 = M_s(\infty,\theta)/L^2 + aL^{-d/2},
$$
in order to obtain the values in the thermodynamic limit.
In this way we estimate the staggered magnetization per spin
at the algorithmic transition point, i.e., $\theta = -\pi/2$ as
\begin{equation}
  M_s(\theta=-\pi/2)/L^2 \simeq 0.14. \label{eq:StaggeredAtTransition}
\end{equation}
This value is small but finite, whereas for $\theta = -0.505\pi$,
we find the staggered magnetization is vanishing.
Therefore, one of the following two possibilities is correct.
One possibility is that
the antiferromagnetic-paramagnetic transition is of the second order
and very close to but slightly below $-\pi/2$, and the other is
that the transition is of the first order and occurs somewhere
in the interval $(-0.505\pi, -0.500\pi]$ including
$\theta = -\pi/2$.

\deffig{fig:StaggeredII}{6cm}{
Staggered magnetization per spin in the thermodynamic limit.
}{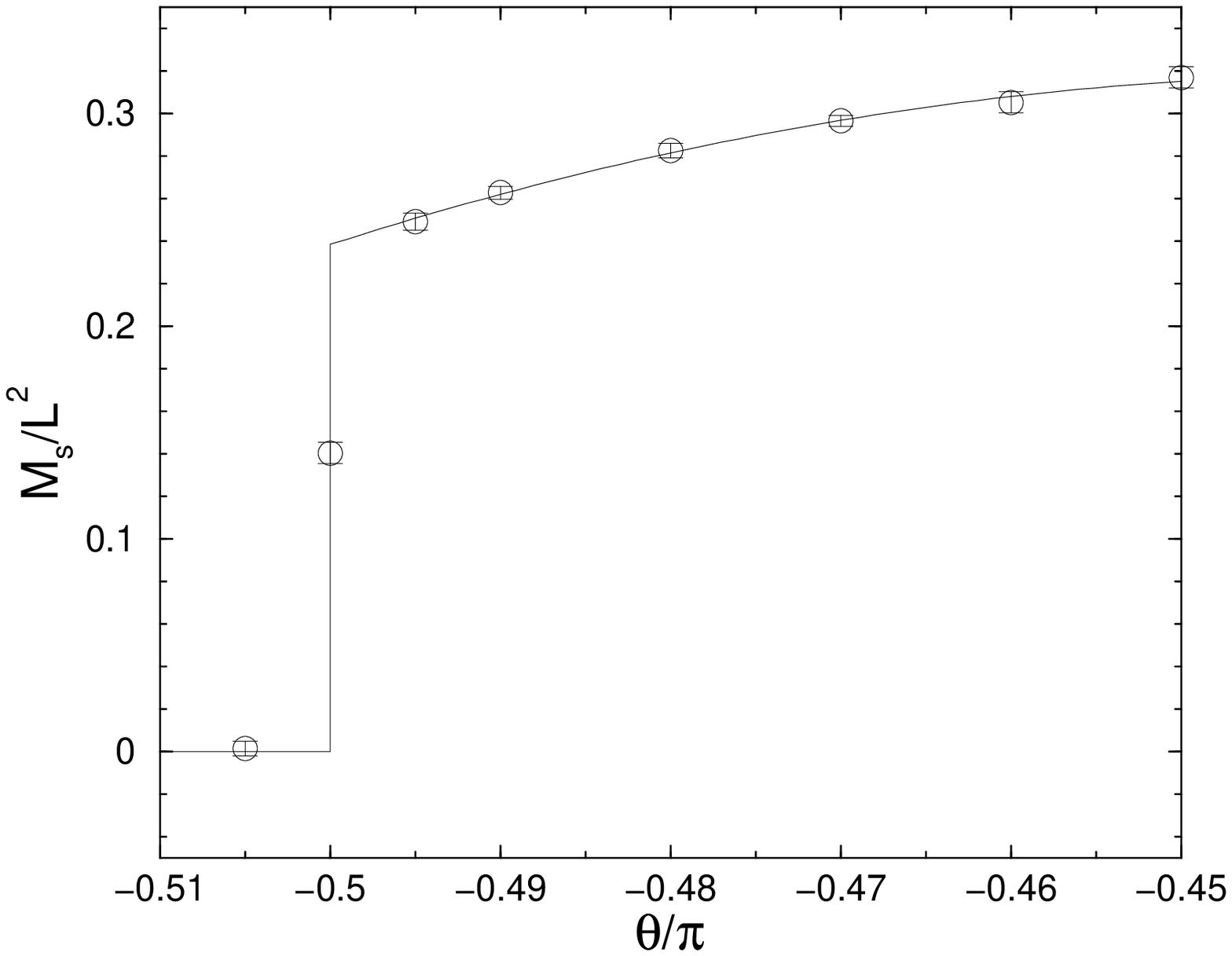}{0.38}

In \Fig{fig:StaggeredII}, we plot $M_s/L^2$ in the
thermodynamic limit (extrapolated first to $\beta =\infty$ and then
to $L =\infty$) as a function of $\theta$.
It seems that, as one approaches this point from above,
$M_s/L^2$ converges to a finite value different from
the one in \Eq{eq:StaggeredAtTransition}.
We estimate this limiting value as
\begin{equation}
  \lim_{\theta\searrow -\pi/2} M_s(\theta)/L^2 \simeq 0.24.
\end{equation}
This suggests that the latter of the two possibilities
is correct and that the transition is exactly at the
algorithmic transition point.

Regarding the intermediate phase $-3\pi/4 < \theta < -\pi/2$, 
we can see no evidence of any kind of order.
At least we see that there is no long-range correlation in
the energy, the magnetization, the staggered magnetization,
or the $(\pi,0)$ or $(\pi,\pi)$ structure factor of the dimer
order parameter.


To summarize, we have presented a new loop algorithm for the Heisenberg 
model with the biquadratic term.
In the case of $S=1$ it covers the entire region of $\theta < 0$,
while it covers a part of it for $S>1$.
Outside these regions, not only the loop algorithm but also the general 
world-line Monte Carlo simulation suffers from the negative sign difficulty.
The algorithmic transition points have been found to be special
from the physical point of view.
In the two-dimensional case, in particular, they correspond to the
real transition points.
In contrast to the one-dimensional case, the intermediate phase is disordered.
A natural interpretation of the present results also suggests that
the transition into this phase from the antiferromagnetic phase is
of the first order.
The extension of the algorithm to the cases where the simple methods
encounter the negative signs is an open problem.


We thank H.-P.~Ying for drawing our attention to the present problem.
The present work is financially supported by
Grant-in-Aid for Scientific Research Programs (No.11740232 and No.12740232) 
from the Ministry of Education, Science, Sports and Culture of Japan.

\end{document}